\documentclass[12pt]{article}
\usepackage{subfloat}
\pdfoutput = 1
\begin{document}
\date{Today}
\title{{\bf{\Large  Emergent Universe with particle production }}}

\author{
{\bf {\normalsize Sunandan Gangopadhyay}$^{a,b}
$\thanks{sunandan.gangopadhyay@gmail.com, sunandan@associates.iucaa.in}},
{\bf {\normalsize Anirban Saha}
$^{a,b}$\thanks{anirban@associates.iucaa.in}}\\
$^{a}$ {\normalsize Department of Physics, West Bengal State University, Barasat, India}\\
$^{b}${\normalsize Visiting Associate in Inter University Centre for Astronomy \& Astrophysics,}\\
{\normalsize Pune, India}\\[0.1cm]
{\bf {\normalsize S. Mukherjee}$^{c,}
$\thanks{sailom47@rediffmail.com}}\\
$^{c}${\normalsize Inter University Centre for Astronomy \& Astrophysics Reference Centre}\\
{\normalsize Kolkata, India}\\[0.3cm]
}
\date{}

\maketitle

\begin{abstract}
{\noindent The possibility of an emergent universe solution to Einstein's field equations
allowing for an irreversible creation of matter at the expense of the gravitational field is shown. 
With the universe being chosen as spatially flat FRW spacetime together with 
equation of state proposed in \cite{sailom}, the solution exists 
when the ratio of the phenomenological matter creation rate to the number density
times the Hubble parameter is a number $\beta$ of the order of unity and independent of time. The thermodynamic
behaviour is also determined for this solution. Interestingly, we also find that an emergent universe scenario is present
with usual equation of state in cosmology when the matter creation rate is chosen to be a constant.
More general class of emergent universe solutions are also discussed.
  }
\end{abstract}
\vskip 1cm

\section{Introduction  }

The origin of cosmological entropy and material content filling the presently
observed universe is an outstanding issue in cosmology. There has been a lot 
of investigations to uncover the mystery behind the creation of matter \cite{parker}-\cite{limaprd}. One of the
early attempts was made by Parker and coworkers\cite{parker, parker1}, where the possibility of the production
of matter quantum mechanically in the context of Einstein's general theory of relativity was shown.
Another interesting approach was developed independently by Tyron \cite{tyron} and Fomin \cite{fomin}. Their
hypothesis was that the universe has been created as a vacuum fluctuation if one assumes that the net
value of all conserved quantities (for example total energy) of the universe is zero.
Brout, Englert and Gunzig, later on proposed a model that could account for a simultaneous
generation of matter and curvature from a quantum fluctuation of the Minkowski spacetime vacuum \cite{brout}.
However, although these models could account for matter creation, they failed to provide the
explanation for the production of entropy accompanying the process of creation of matter.
The problem lies in the fact that Einstein's classical field equations and even semiclassical Einstein's equations 
are purely adiabatic and reversible and therefore fail to account for entropy production
during matter creation.

The question of incorporating entropy burst along with creation of matter in the framework of Einstein's field equations
was considered in depth by Prigogine et.al \cite{prigogine}. They argued that production of matter can occur only as an irreversible process at the expense of the gravitational field. The important aspect of this formalism is the introduction
of a balance equation for the number of created particles along with the field equations of Einstein. The combination
of this equation with the second law of thermodynamics yields an additional negative pressure term which depends 
on the rate of matter creation. These results were further generalized in \cite{limapla} where provision for variation of specific entropy was made through a covariant formulation.

The search for a singularity free inflationary model in cosmology has been another prominent issue in this field. 
Recently, to avoid the timelike singularity, investigations has been carried out to look for a universe which is ever existing and large enough so that the spacetime may be treated classically. This model is usually called an emergent universe and has an almost static behaviour in the infinite past ($t\rightarrow-\infty$) and evolves eventually into an inflationary stage \cite{kn:4}-\cite{kn:2}. 
Various models of emergent universe have been proposed in the literature.
The most prominent ones are that of a closed universe containing radiation and a cosmological constant \cite{kn:4}
and a closed universe containing a minimally coupled scalar field $\phi$ having a self-interaction $V(\phi)$ \cite{kn:1}.
Another interesting approach to arrive at an emergent universe scenario has been given in \cite{sailom}. Here, a mixture of matter and exotic matter are prescribed by an equation of state (EOS). 
The analysis has been carried out in a flat universe which is in accordance with present cosmological observations. The emergent universe
scenario in the presence of interacting fluids have also been studied recently \cite{archan}.
This picture have received considerable support
from other approaches too. An approach based on the holographic principle have been provided to study the emergence of space and cosmology in \cite{paddy}.
Further, analysis involving the deformation of canonical commutation relations consistent with the existence of a minimum length scale have shown that the big bang singularity can be avoided naturally \cite{faizal}-\cite{garattini}.

In this paper, we show that there exists an emergent universe solution to the Einstein's equations which allow for an irreversible production of matter from the gravitational field ala Prigogine et.al \cite{prigogine}. We observe that such a solution exists (with the EOS proposed in \cite{sailom}) when the ratio of the phenomenological matter creation rate to the number density times the Hubble parameter is a number ($\beta$) of the order of unity and independent of time. We also determine the thermodynamic behaviour for the matter creation formulation 
which leads to this emergent universe scenario and compare our results with \cite{limaprd} where the thermodynamic
behaviour was obtained with the standard EOS for a fluid. Finally, we observe that an emergent universe scenario
also exits with usual EOS in cosmology when the matter creation rate is chosen to be a constant. Such a universe
does not have exotic matter unlike the earlier one. More general class of emergent universe solutions are also found
when the constant matter creation rate is augmented by a term proportional to the Hubble parameter.

The paper is organized as follows. In the next section, we obtain emergent universe solutions to the
Einstein's equations allowing for irreversible creation of matter.  We then conclude in section 3.



\section{Emergent universe scenario with matter creation}

\noindent To begin with, we write down the Einstein equations for a flat universe in FRW metric augmented
by the balance equation for matter creation \cite{prigogine}
\begin{eqnarray}
\label{e1} 
3 \frac{ \dot{a}^{2}}{a^{2}}&=&\rho\\
\label{e2}
2\frac{ \ddot{a}}{a} + \frac{ \dot{a}^{2}}{a^{2}}&=&-(p-p_c)\\
\frac{\dot{n}}{n}+3\frac{\dot{a}}{a}&=&\frac{\psi}{n}
\label{e3}
\end{eqnarray}
where
\begin{eqnarray}
\label{e4} 
p_c = \frac{\rho+p}{3nH}\psi
\end{eqnarray}
and $H=\dot{a}/a$ is the Hubble parameter. An overdot means time derivative and $\rho$, $p$, $n$ and $\psi$
are the energy density, thermostatic pressure, particle number density and matter creation rate.
Let us also introduce the dimensionless and in general time-dependent parameter
\begin{eqnarray}
\label{e5} 
\beta(t) = \frac{\psi}{3nH}\quad;\quad\beta\geq0
\end{eqnarray}
in order to measure the effects of the matter creation rate $\psi$.

\noindent In the subsequent discussion, we shall consider two EOS(s). We shall first consider the EOS proposed in \cite{sailom}
\begin{equation}
p(\rho) = A \rho - B \rho^{\frac{1}{2}}
\label{eos}
\end{equation}
where $A$ and $B$ are constants and the energy density $\rho$ may have different components, each satisfying its own equation of state.

\noindent Making use of eq.(s)(\ref{e1}, \ref{e2}, \ref{eos}) and eq.(s)(\ref{e4}, \ref{e5}), we obtain
\begin{eqnarray}
2\frac{ \ddot{a}}{a} + \left\{(1+ 3A)-3(1+A)\beta(t)\right\}\frac{\dot{a}^{2}}{a^{2}}-\sqrt{3}B[1-\beta(t)]\frac{ \dot{a}}{a} = 0.
\label{diffeqn}
\end{eqnarray}
Assuming $\beta$ to be a constant \cite{limaprd}, this equation can be easily integrated to give 
\begin{equation}
a(t) = c_1\left\{c+\frac{\sqrt{3}(1+A)}{B}e^{\frac{\sqrt{3}}{2}B(1-\beta)t}\right\}^{\frac{2}{3 (1+A)(1-\beta)}}
\label{sol}
\end{equation}
where $c$ and $c_1$ are constants of integration. It is important to observe that the solution describes an emergent universe
for $B > 0$,  $(1+A)>0$, $\beta<1$ and $c, c_1 >0$. The restriction on $\beta$ to be time independent and 
less than unity comes out naturally in order to have an 
emergent universe scenario\footnote{Matter creation models with $\beta<1$ has also been 
considered in \cite{limaprd}.}. Also, the above result matches with the result in \cite{sailom} for $\beta=0$ 
which corresponds to no particle production.

\noindent To investigate the possible composition of the emergent universe with particle production, 
we first study the dependence of the energy density on the scale factor. To proceed, we consider the energy conservation equation
(which follows from eq.(s)(\ref{e1}, \ref{e2}))
\begin{equation}
\dot{\rho} + 3 (\rho + p-p_c) \frac{ \dot{a}}{a}  = 0.
\label{energy}
\end{equation}
Making use of eq.(s)(\ref{e4}) and (\ref{eos}), we obtain the following differential equation for $\rho$ :
\begin{equation}
\dot{\rho} + 3\left\{(1+A)\rho-B\rho^{1/2}\right\}(1-\beta)\frac{ \dot{a}}{a}=0.
\label{e10}
\label{energy}
\end{equation}
This can be integrated to give
\begin{equation}
\rho(a)=\frac{ 1}{(1+A)^{2}}\left\{B +  \frac{ K}{a^{3(1+A)(1-\beta)/2}} \right\}^{2}
\label{e11}
\end{equation}
where $K$ is an integration constant. Note that $K$ should be negative and this can be seen readily by computing the
Hubble parameter for the emergent universe solution (\ref{sol}):
\begin{equation}
H=\frac{e^{\frac{\sqrt{3}}{2}B(1-\beta)t}}{c+\frac{\sqrt{3}(1+A)}{B}e^{\frac{\sqrt{3}}{2}B(1-\beta)t}}~.
\label{hubble}
\end{equation}
The above expression goes to zero in the limit $t\rightarrow-\infty$ and therefore (from eq.(\ref{e1})) the energy density $\rho$ also
vanishes in this limit. This implies that $K$ must be negative.

\noindent One can also write $\rho$ as a function of $n$ by writing eq.(\ref{energy}) in the following form
\begin{equation}
\frac{\dot{\rho}}{(1+A)\rho-B\rho^{1/2}}+3\frac{\dot{a}}{a}=\frac{\psi}{n}~.
\label{e11a}
\end{equation}
Comparing the left hand sides of this equation and eq.(\ref{e3}), we get
\begin{equation}
\frac{\dot{\rho}}{(1+A)\rho-B\rho^{1/2}}+3\frac{\dot{a}}{a}=\frac{\dot{n}}{n}
\label{e12}
\end{equation}
which can be integrated to give
\begin{equation}
\rho(n)=\frac{ 1}{(1+A)^{2}}\left\{B+\frac{n^{(1+A)/2}}{K'}\right\}^{2}
\label{e13}
\end{equation}
where $K'$ is an integration constant.

\noindent An equation connecting $n$ as a function of $a$ can also be obtained by solving eq.(\ref{e3}): 
\begin{equation}
n(a)=\frac{c_2}{a^{3(1-\beta)}}
\label{e130}
\end{equation}
where $c_2$ is an integration constant.

\noindent Eq.(\ref{e11}) provides us the information about the components of energy density that lead to an emergent universe.  
We can rewrite this equation as
\begin{eqnarray}
\rho=\frac{ B^{2}}{(1+A)^{2}} +  \frac{2 K B}{(1+A)^{2}}  
\frac{ 1}{a^{3(1+A)(1-\beta)/2}} + \frac{K^{2}}{(1+A)^{2}}  
\frac{1}{a^{3(1+A)(1-\beta)}}=\rho_{1} + \rho_{2} + \rho_{3}~.
\label{e14}
\end{eqnarray}
Substituting this expression in eq.(\ref{eos}), we obtain the pressure $p$ to be
\begin{equation}
p=-\frac{ B^{2}}{(1+A)^{2}} +  \frac{ KB(A-1)}{(1+A)^{2}}  
\frac{ 1}{ a^{3(1+A)(1-\beta)/2}} + \frac{  A K^{2}}{(1+A)^{2}}  
\frac{ 1}{ a^{3(1+A)(1-\beta)}}=p_{1} + p_{2} + p_{3}~.
\label{e15}
\end{equation}
Eq.(s)(\ref{e14}, \ref{e15}) has the same form as the corresponding ones in \cite{sailom} except the fact that the matter creation
rate measured by the parameter $\beta$ appears in these equations explicitly. This in turn implies that although the composition 
of the universe remains the same as in \cite{sailom}, $\rho_2$ scales with $a^{-3(1+A)(1-\beta)/2}$ and 
$\rho_3$ scales with $a^{-3(1+A)(1-\beta)}$ showing once again the role played by matter creation. 
The details of the composition of the universe are presented in an appendix.

\noindent We now determine the thermodynamic behaviour of the matter creation formalism with the EOS (\ref{eos})
which leads to the emergent universe solution. To proceed, we write down the equation for the
rate of variation of temperature $T$ which reads \cite{limapla}
\begin{equation}
\frac{\dot{T}}{T}=\left(\frac{\partial p}{\partial\rho}\right)_{n}\frac{\dot{n}}{n}~.
\label{t1}
\end{equation}
Using the EOS (\ref{eos}) and eq.(\ref{e13}), the above equation takes the form
\begin{eqnarray}
\frac{\dot{T}}{T}=\left\{A-\frac{(1+A)BK'}{2(BK' +n^{(1+A)/2})}\right\}\frac{\dot{n}}{n}~.
\label{t2}
\end{eqnarray}
This can be readily integrated to give
\begin{eqnarray}
\frac{T}{T_0}=\left(\frac{n_0}{n}\right)^{(1-A)/2}\frac{\{BK' + n^{(1+A)/2}\}}{\{BK' + n_{0}^{(1+A)/2}\}}~.
\label{t3}
\end{eqnarray}
Hence the relation between the temperature and number density is quite different than the one which uses the
standard ``gamma-law" EOS $p=(\gamma-1)\rho$ \cite{limaprd}. It is easy to see that 
the dependence (between temperature and number density) 
in \cite{limaprd} can be recovered from the above result by putting $B=0$.

\noindent We can also obtain the relation between the temperature and scale factor from eq.(s)(\ref{t3}, \ref{e130}):
\begin{eqnarray}
\frac{T}{T_0}=\left(\frac{a_0}{a}\right)^{3A(1-\beta)}
\frac{\{BK' a^{3(1+A)(1-\beta)/2}+1\}}{\{BK' a_{0}^{3(1+A)(1-\beta)/2}+1\}}~.
\label{t4}
\end{eqnarray}
This relation also differs from the temperature law given in \cite{limaprd}.

\noindent We now carry out the investigation using the usual EOS employed in cosmology
\begin{eqnarray}
p=A\rho~.
\label{usualeos}
\end{eqnarray}
Once again making use of eq.(s)(\ref{e1}, \ref{e2}, \ref{usualeos}) and eq.(s)(\ref{e4}, \ref{e5}), we obtain
\begin{eqnarray}
2\frac{ \ddot{a}}{a} + \left\{(1+ 3A)-3(1+A)\beta\right\}\frac{\dot{a}^{2}}{a^{2}} = 0.
\label{u1}
\end{eqnarray}
Choosing the matter creation rate $\psi$ to be of the form \cite{pavon}
\begin{eqnarray}
\psi=3n\zeta\quad;\quad\zeta>0
\label{u2}
\end{eqnarray}
where $\zeta$ is a constant, fixes $\beta$ (using eq.(\ref{e5})) to be
\begin{eqnarray}
\beta=\frac{\zeta}{H}~.
\label{u2a}
\end{eqnarray}
The above equation can then be integrated to give
\begin{equation}
a(t) = e''\left\{e'+\frac{1}{\zeta}e^{\frac{3}{2}(1+A)\zeta t}\right\}^{\frac{2}{3 (1+A)}}
\label{u3}
\end{equation}
where $e'$ and $e''$ are constants of integration. It is important to observe that the solution describes an emergent universe for $(1+A)>0$ and $e', e'' >0$. The above analysis reveals that one can obtain an emergent universe solution
in the matter creation formalism with usual composition of matter (i.e. without exotic matter). 

\noindent As before, one can once again work out the relation between $\rho$ and $a$.
To do this, we use eq.(s)(\ref{energy}, \ref{e4}) and (\ref{usualeos}) to obtain the following differential equation for $\rho$ :
\begin{equation}
\dot{\rho} + 3(1+A)\rho\frac{ \dot{a}}{a}=3(1+A)\zeta\rho.
\label{u4}
\end{equation}
Integrating this equation yields
\begin{equation}
\rho(a)=\frac{c'  e^{3(1+A)\zeta t} }{a^{3(1+A)}}
\label{u5}
\end{equation}
where $c'$ is an integration constant. Solving eq.(\ref{e3}) with $\psi$ given by eq.(\ref{u2}) gives
\begin{equation}
n(a)=c''\frac{e^{3\zeta t}}{a^{3}}
\label{u6}
\end{equation}
where $c''$ is an integration constant. This in turn (using eq.(\ref{u5})) leads to the following relation between $\rho$ and $n$ :
\begin{equation}
\rho(n)=k' n^{1+A}
\label{u7}
\end{equation}
where $k'$ is a constant. Note that eq(s)(\ref{u5}, \ref{u6}, \ref{u7}) 
are quite different in form than eq(s)(\ref{e11}, \ref{e130}, \ref{e13}) obtained using the EOS (\ref{eos}).

\noindent It is also possible to obtain more general class of emergent universe solutions by
choosing the matter creation rate $\psi$ to be of the form
\begin{eqnarray}
\psi=3n(\zeta+\alpha H)\quad;\quad\alpha\geq0~.
\label{u8}
\end{eqnarray}
This fixes $\beta$ (using eq.(\ref{e5})) to be
\begin{eqnarray}
\beta=\frac{\zeta}{H}+\alpha~.
\label{u2aa}
\end{eqnarray}
Eq.(\ref{u1}) can then be integrated to give
\begin{equation}
a(t) = e''\left\{e'+\frac{(1-\alpha)}{\zeta}e^{\frac{3}{2}(1+A)\zeta t}\right\}^{\frac{2}{3 (1+A)(1-\alpha)}}~.
\label{u3}
\end{equation}
The solution describes an emergent universe for $(1+A)>0$, $\alpha<1$, $\zeta>0$ and $e', e'' >0$.


\section{Conclusions}

In this paper, we have shown the existence of an emergent universe scenario
to the matter creation formulation of Prigogine et.al \cite{prigogine} in the framework of Einstein's equations.
The solution is therefore compatible with the production of entropy accompanying the matter creation process at the
expense of the gravitational field. We analyse two cases. Considering the EOS proposed in \cite{sailom}, we find
that the emergent universe scenario exists when the ratio of the phenomenological matter creation rate to the number density
times the Hubble parameter is a number $\beta$ less than unity and independent of time. 
The physical content of the universe from this solution remains the same as in
\cite{sailom} except that the parameter 
$\beta$ (which is a measure of the matter creation rate) appears in the scaling relations between the
energy density and the scale parameter. The thermodynamic
behaviour for the matter creation formulation with this EOS is also determined.
We then consider the usual EOS in cosmology. Interestingly, such an EOS admits an emergent universe solution
of Einstein's equations in the matter creation formalism when the matter creation rate is chosen to be a 
constant. The composition of this universe in this case does not have exotic matter unlike the one obtained using the EOS
proposed in \cite{sailom}. We also discuss more general class of emergent universe solutions with the usual EOS.
As a future work, it would be interesting to extend our analysis in the
presence of interacting fluids.

\section*{Appendix}
We present two entries in the table below to identify the components from eq(s)(\ref{e14}, \ref{e15}) for $\rho$ and $p$.
The first term in these equations behaves like a cosmological constant and may account for dark energy.
The physical content of the other two terms are listed in the table.

\begin{table}[htbp]
\centerline{\footnotesize Table 1.}
\begin{tabular}{l c c c c c c} \\
\hline \\
A   &  $\frac{\rho_{2}}{ \Lambda }$ in unit $\frac{ K}{B}$ & $\omega_{2}=\frac{A-1}{2}$ & 
$ \frac{\rho_{3}}{ \Lambda}  $ in unit $(\frac{ K}{B})^{2}$  & $\omega_{3}=A$ & Composition \\
\\
\hline \\
$ \frac{1}{3}$ &\phantom0$\frac{9}{8a^{2(1-\beta)}}$ & $-\frac{1}{3}$ & 
$\frac{9}{16a^{4(1-\beta)}}$ & $ \frac{1}{3}$& dark energy, \\
{} & & & &  &  exotic matter and radiation \\
\hline \\
$-\frac{1}{3}$ &\phantom0$\frac{9}{2a^{1-\beta}}$&   $-\frac{2}{3}$ & 
$\frac{9}{4a^{2(1-\beta)}}$ & $-\frac{1}{3}$& dark energy \\
{} & & & &  &  exotic matter and cosmic strings  \\
\hline \\
\end{tabular}
\caption{ \it  Composition of universal matter for various   values of A }
\end{table}

\noindent We observe that the matter creation rate measured by the parameter $\beta$ modifies the
exponent of the scale factor $a$.


\vspace{0.5 cm}

\noindent {\bf{Acknowledgements}}: S.G. acknowledges the support by DST SERB under Start Up Research Grant (Young Scientist), File No.YSS/2014/000180. 
AS acknowledges the financial support of DST SERB under
Grant No. SR/FTP/PS-208/2012.
The authors also thank the referee for useful comments.


\end{document}